\documentclass[aps,prd,twocolumn,superscriptaddress,showpacs,showkeys,
nofootinbib]{revtex4-1}

\usepackage{latexsym}
\usepackage{amsmath}
\usepackage{amssymb}
\usepackage{amsfonts}

\usepackage{color}

\usepackage{supertabular} 
\usepackage{placeins}
\usepackage{epsfig}
\usepackage{graphicx}

\usepackage{epstopdf}

\begin{document}


\title{Canonical description of the new LHCb resonances}



\author{Pablo G. Ortega}
\email[]{pgortega@ific.uv.es}
\affiliation{Instituto de F\'isica Corpuscular (IFIC), 
Centro Mixto CSIC-Universidad de Valencia, ES-46071 Valencia, Spain}

\author{Jorge Segovia}
\email[]{jorge.segovia@tum.de}
\affiliation{Physik-Department, Technische Universit\"at M\"unchen, 
James-Franck-Str.~1, 85748 Garching, Germany}

\author{David R. Entem}
\email[]{entem@usal.es}

\author{Francisco Fern\'andez}
\email[]{fdz@usal.es}
\affiliation{Grupo de F\'isica Nuclear and Instituto Universitario de F\'isica 
Fundamental y Matem\'aticas (IUFFyM), Universidad de Salamanca, E-37008 
Salamanca, Spain}


\date{\today}

\begin{abstract}
The LHCb Collaboration has recently observed four $J/\psi\phi$ structures called 
$X(4140)$, $X(4274)$, $X(4500)$ and $X(4700)$ in the $B^{+}\to J/\psi\phi K^{+}$ 
decays. We study them herein using a nonrelativistic constituent quark 
model in which the degrees of freedom are quark-antiquark and meson-meson 
components. The $X(4140)$ resonance appears as a cusp in the $J/\psi\phi$ 
channel due to the near coincidence of the $D_{s}^{\pm}D_{s}^{\ast\pm}$ and 
$J/\psi\phi$ mass thresholds. The remaining three $X(4274)$, $X(4500)$ and 
$X(4700)$ appear as conventional charmonium states with quantum numbers 
$3^{3}P_{1}$, $4^{3}P_{0}$ and $5^{3}P_{0}$, respectively; and whose masses 
and widths are slightly modified due to their coupling with the corresponding 
closest meson-meson thresholds. A particular feature of our quark model is a 
lattice-based screened linear confining interaction that has been constrained 
in the light quark sector and usually produces higher excited heavy-quark 
states with lower masses than standard quark model predictions.
\end{abstract}



\maketitle

One of the basic properties of Quantum Chromodynamics (QCD) is its spectrum: 
the list of particles that are stable or at least sufficiently long-lived to be 
observed as resonances. The elementary constituents in QCD are quarks $(q)$, 
antiquarks $(\bar{q})$, and gluons $(g)$, and QCD requires them to be confined 
into colour-singlet clusters called hadrons. The most stable hadrons are the 
clusters predicted by the quark model~\cite{GellMann:1964nj, Zweig:1964CERN}, 
conventional mesons $(q\bar{q})$, baryons $(qqq)$ and antibaryons $(\bar q\bar 
q\bar q)$, which have been the only states observed in experiments for around 
$30$ years~\cite{Agashe:2014kda}.

This simple picture is being challenged since $2003$ with the discovery of 
almost two dozen charmonium- and bottomonium-like XYZ states that do not fit the 
naive quark-antiquark interpretation. Most of these states usually appear close 
to meson-meson thresholds and thus their dynamics can be strongly dictated by 
the nearby multiquark channels. In fact, the discovery of the XYZ particles is 
opening the door to systematically explore higher Fock components of the meson 
wave function such as molecules, compact tetraquarks or diquark-antidiquark 
(diquarkonium) structures. 

The $X(4140)$, $X(4274)$, $X(4500)$ and $X(4700)$ are some of the last 
XYZ states observed, this time, in the amplitude analysis of $B^{+}\to 
J/\psi\phi K^{+}$ decays performed by the LHCb 
Collaboration~\cite{Aaij:2016iza, Aaij:2016nsc}. The $X(4140)$ was previously 
seen by CDF~\cite{Aaltonen:2011at}, D0~\cite{Abazov:2015sxa}, 
CMS~\cite{Chatrchyan:2013dma}, Belle~\cite{Shen:2009vs} and 
BaBar~\cite{Lees:2014lra} Collaborations; the rest: $X(4274)$, $X(4500)$ and 
$X(4700)$ have been determined for the first time with the LHCb experiment.

Thanks to the large signal yield, the roughly uniform efficiency and the 
relatively low background across the entire $J/\psi \phi$ mass range, the LHCb 
data~\cite{Aaij:2016iza, Aaij:2016nsc} offers the best sensitivity to date in 
order to probe the nature of the observed structures, but also to establish 
their quantum numbers. The quantum numbers of the $X(4140)$ and $X(4274)$ are 
determined to be $J^{PC}=1^{++}$ with statistical significance $5.7\sigma$ and 
$5.8\sigma$, respectively. The $X(4500)$ and $X(4700)$ resonances have both 
$J^{PC}=0^{++}$ with statistical significance $4.0\sigma$ and $4.5\sigma$, 
respectively. 

The determination by the LHCb Collaboration of the $J^{PC}=1^{++}$ quantum 
numbers for the $X(4140)$ and $X(4274)$ resonances have had a big impact on 
their theoretical interpretations, ruling out most of the multiquark models. 
Lebed-Polosa~\cite{Lebed:2016yvr} predicted the $X(4140)$ to be a $1^{++}$ 
tetraquark but they expected the $X(4274)$ peak to be $0^{-+}$ in the same 
model. Molecular interpretations~\cite{Liu:2009ei, Branz:2009yt, 
Albuquerque:2009ak, Ding:2009vd, Zhang:2009st} found that the $X(4140)$ can only 
be a $J^{PC}=0^{++}$ or $2^{++}$ $D_{s}^{\ast+}D_{s}^{\ast-}$ molecule. Compact 
tetraquark models expected $0^{-+},\,1^{-+}$~\cite{Drenska:2009cd}, or 
$0^{++},\,2^{++}$~\cite{Wang:2015pea} state(s) in the mass region of interest. 
Finally, no evidence of a $1^{++}$ tetraquark state has been found in a 
lattice-regularised QCD computation with diquark 
operators~\cite{Padmanath:2015era}.

A similar situation can be found for the $J^{PC}=0^{++}$ $X(4500)$ and 
$X(4700)$ resonances discovered in the high $J/\psi\phi$ mass region. For 
instance, the work of Wang {\it et al.}~\cite{Wang:2009ry} predicted only 
one virtual $D_{s}^{\ast+}D_{s}^{\ast-}$ state at $4.48\pm0.17\,{\rm GeV}$. 
Therefore, the novelty of these states resides in the fact that it is difficult 
to explain their structure as molecules or tetraquarks.\footnote{It is fair to 
mention that the $X(4274)$ meson appears as a natural quark-gluon hybrid 
candidate since the lowest spin-multiplet predicted by Lattice 
QCD~\cite{Liu:2012ze} has an average mass of $4281\pm16\,{\rm MeV}$. However, 
the quantum numbers of the $X(4274)$ has been established to be $1^{++}$ and so 
this state cannot belong to that multiplet but to the higher one with an average 
mass of $4383\pm30\,{\rm MeV}$.}
 
In this situation, we first have to remember that coupled-channel effects can 
generate signals which mimic resonances. These structures may appear near 
two-particle thresholds if the attraction between the two particles in the 
channel is not sufficient to produce a resonance but the amplitude behaves as it 
would be a resonance (cusp)~\cite{Bugg:2008kd}. The near coincidence of the 
$D_{s}^{\pm}D_{s}^{\ast\pm}$ and $J/\psi\phi$ mass thresholds provides suitable 
conditions to form cusps. An extensive study of the possible rescattering 
effects which may contribute to the process $B^{+}\to J/\psi\phi K^{+}$ has 
been performed in, for instance, Ref.~\cite{Liu:2016onn}.

One should also not forget that, in general, a meson can be compose of 
multiquark Fock components as $|M\rangle = |q\bar q\rangle + |q q\bar q\bar 
q\rangle + \ldots$ but the dominant Fock space component is the $q\bar q$ one 
and thus pure (or dominant) higher excited quark-antiquark states predicted by 
the naive quark model can appear in the experimental measurements when 
exploring the higher energy spectrum.

In this work we shall show that the $X(4140)$ can be interpreted as a cusp in 
the $J/\psi\phi$ channel due to the presence of the $D_{s}D_{s}^{\ast}$ 
threshold, whereas the rest of states observed by the LHCb Collaboration in the 
$J/\psi\phi$ invariant mass correspond to quark-antiquark structures whose 
mass is slightly renormalized by the presence of nearby meson-meson thresholds.


We use the constituent quark model (CQM) presented in~\cite{Vijande:2004he} and 
updated in~\cite{Segovia:2008zz} (see Refs.~\cite{Valcarce:2005em} 
and~\cite{Segovia:2013wma} for reviews). The CQM is based on the assumption 
that the light-quark constituent mass appears owing to the dynamical breaking 
of chiral symmetry in QCD at some momentum scale. Regardless of the breaking 
mechanism, the simplest Lagrangian which describes this situation must contain 
chiral fields to compensate the mass term and can be expressed 
as~\cite{Diakonov:2002fq}
\begin{equation}
{\mathcal L} = \bar{\psi}(i\, {\slash\!\!\! \partial} 
-M(q^{2})U^{\gamma_{5}})\,\psi  \,,
\end{equation}
where $U^{\gamma_{5}}=\exp(i\pi^{a}\lambda^{a}\gamma_{5}/f_{\pi})$, $\pi^{a}$ 
denotes nine pseudoscalar fields $(\eta_{0},\,\vec{\pi },\,K_{i},\,\eta _{8})$ 
with $i=1,\ldots,4$ and $M(q^2)$ is the constituent mass. This constituent 
quark mass, which vanishes at large momenta and is frozen at low 
momenta at a value around $350\,{\rm MeV}$, can be explicitly obtained from the 
underlying theory but its theoretical point wise behaviour is simulated herein 
by parameterizing $M(q^{2}) = m_{q}F(q^{2})$ with $m_{q}$ the bare quark 
mass and
\begin{equation}
F(q^{2}) = \left[\frac{{\Lambda}^{2}}{\Lambda^{2}+q^{2}}\right]^{\frac{1}{2}} 
\,,
\end{equation} 
where the cut-off $\Lambda$ fixes the chiral symmetry breaking scale.

The matrix of Goldstone-boson fields can be expanded in the following form
\begin{equation}
U^{\gamma _{5}} = 1 + \frac{i}{f_{\pi }} \gamma^{5} \lambda^{a} 
\pi^{a} - \frac{1}{2f_{\pi }^{2}} \pi^{a} \pi^{a} + \ldots
\end{equation}
The first term of the expansion generates the constituent quark mass while the
second gives rise to a one-boson exchange interaction between quarks. The
main contribution of the third term comes from the two-pion exchange 
interaction which has been simulated by means of a scalar exchange potential.

In the heavy quark sector chiral symmetry is explicitly broken and we do not 
need to introduce additional fields. However, the chiral fields introduced 
above provide a natural way to incorporate the pion-exchange interaction in the 
molecular dynamics.

The next ingredient of our quark model is the non-relativistic limit of 
one-gluon exchange (OGE) interaction, the Breit-Fermi interaction, in analogy 
to positronium. The OGE potential is generated from the vertex 
Lagrangian~\cite{DeRujula:1975qlm}
\begin{equation}
{\mathcal L}_{qqg} = i\sqrt{4\pi\alpha_{s}} \,\, \bar{\psi}\, 
\gamma_{\mu}\, G^{\mu}_{c}\, \lambda^{c}\, \psi \,,
\label{Lqqg}
\end{equation}
where $\lambda^{c}$ are the $SU(3)$ colour matrices, $G^{\mu}_{c}$ is the
gluon field and $\alpha_{s}$ is the strong coupling constant. The scale 
dependence of $\alpha_{s}$ can be found in {\it e.g.} 
Ref.~\cite{Vijande:2004he}, it allows a consistent description of light, strange 
and heavy mesons.

The last main feature of our constituent quark model is based on the empirical 
fact that quarks and gluons have never seen as isolated particles. Colour 
confinement should be encoded in the non-Abelian character of QCD, however, at 
present, it is still infeasible to analytically derive these property from the 
QCD Lagrangian. Lattice-regularised QCD studies have demonstrated 
that multi-gluon exchanges produce an attractive linearly rising potential 
proportional to the distance between infinite-heavy quarks~\cite{Bali:2005fu}. 
However, the spontaneous creation of light-quark pairs from the QCD vacuum may 
give rise at the same scale to a breakup of the colour 
flux-tube~\cite{Bali:2005fu}. We have tried to mimic these two phenomenological 
observations by the expression:
\begin{equation}
V_{\rm CON}(\vec{r}\,)=\left[-a_{c}(1-e^{-\mu_{c}r})+\Delta \right] 
(\vec{\lambda}_{q}^{c}\cdot\vec{\lambda}_{\bar{q}}^{c}) \,,
\label{eq:conf}
\end{equation}
where $a_{c}$ and $\mu_{c}$ are model parameters. At short distances this 
potential presents a linear behaviour with an effective confinement strength, 
while at large distances it shows a threshold from which no quark-antiquark 
bound states can be found. This form of the confining potential, slightly 
different from the usual one which grows linearly until infinity, is important 
to describe the higher excited states in the quarkonium spectrum, in 
particular, the charmonium one.

Explicit expressions for all the potentials and the value of the model 
parameters can be found in Ref.~\cite{Vijande:2004he}, updated
in Ref.~\cite{Segovia:2008zz}.

In order to find the quark-antiquark bound states with this constituent quark 
model, we solve the Schr\"odinger equation using the Gaussian expansion 
method~\cite{Hiyama:2003cu} (GEM), expanding the radial wave function in terms 
of basis functions 
\begin{equation}
R_{\alpha}(r)=\sum_{n=1}^{n_{max}} c_{n}^\alpha \phi^G_{nl}(r),
\end{equation} 
where $\alpha$ refers to the channel quantum numbers and $\phi^G_{nl}(r)$ are 
Gaussian trial functions with ranges in geometric progression. This choice is 
useful for optimizing the ranges with a small number of free 
parameters~\cite{Hiyama:2003cu}. In addition, the geometric progression is 
dense at short distances, so that it enables the description of
the dynamics mediated by short range potentials. 

The coefficients, $c_{n}^\alpha$, and the eigenvalue, $E$, are determined from 
the Rayleigh-Ritz variational principle
\begin{equation}
\sum_{n=1}^{n_{max}} \left[\left(T_{n'n}^\alpha-EN_{n'n}^\alpha\right)
c_{n}^\alpha+\sum_{\alpha'}
\ V_{n'n}^{\alpha\alpha'}c_{n}^{\alpha'}=0\right],
\end{equation}
where $T_{n'n}^\alpha$, $N_{n'n}^\alpha$ and $V_{n'n}^{\alpha\alpha'}$ are the 
matrix elements of the kinetic energy, the normalization and the potential, 
respectively. $T_{n'n}^\alpha$ and $N_{n'n}^\alpha$ are diagonal, whereas the
mixing between different channels is given by $V_{n'n}^{\alpha\alpha'}$.


\begin{table}[!t]
\begin{center}
\begin{tabular}{ccccc}
\hline
\hline
State & $J^{PC}$ & $nL$  & Theory (MeV) & Experiment (MeV) \\
\hline
$\chi_{c0}$ & $0^{++}$ &  $3P$ & $4241.7$  & \\
            &          &  $4P$ & $4497.2$  & $4506\pm 11^{+12}_{-15}$ \\
            &          &  $5P$ & $4697.6$  & $4704\pm 10^{+14}_{-24}$ \\
\hline
$\chi_{c1}$ & $1^{++}$ &  $3P$ & $4271.5$  &  $4273.3\pm 8.3$ \\
            &          &  $4P$ & $4520.8$  &                  \\
            &          &  $5P$ & $4716.4$  &                  \\           
\hline
\hline
\end{tabular}
\caption{\label{tab:qqbar} Naive quark-antiquark spectrum in the region of 
interest of the LHCb~\cite{Aaij:2016iza, Aaij:2016nsc} for the $0^{++}$ and 
$1^{++}$ channels.}
\end{center}
\end{table}

Table~\ref{tab:qqbar} shows the calculated naive quark-antiquark spectrum in 
the region of interest of the LHCb for the $J^{PC}=0^{++}$ and $1^{++}$ 
channels. A tentative assignment of the theoretical states with the 
experimentally observed mesons at the LHCb experiment is also given. It can be 
seen that the naive quark model is able to reproduce all the new LHCb resonances 
except the $X(4140)$. The $X(4274)$, $X(4500)$ and $X(4700)$ appear as 
conventional charmonium states with quantum numbers $3^{3}P_{1}$, $4^{3}P_{0}$ 
and $5^{3}P_{0}$, respectively.

Tables~\ref{tab:chic13P},~\ref{tab:chic04P} and~\ref{tab:chic05P} show, 
respectively, the partial and total decay widths of the $X(4274)$, $X(4500)$ and 
$X(4700)$ mesons assuming the above assignment of their quantum numbers. The 
decay widths have been computed using a modified version of the $^{3}P_{0}$ 
decay model presented in Ref.~\cite{Segovia:2012cd}. In such a version, the 
strength $\gamma$ of the decay interaction depends on the mass scale as
\begin{equation}
\gamma(\mu) = \frac{\gamma_{0}}{\log\left(\frac{\mu}{\mu_{0}}\right)},
\label{eq:fitgamma}
\end{equation}
where $\mu$ is the reduced mass of the quark-antiquark in the decaying meson
and, $\gamma_{0}=0.81\pm0.02$ and $\mu_{0}=(49.84\pm2.58)\,{\rm MeV}$ are
parameters determined by the global fit. The value of the $\gamma$ in the 
charmonium sector is $0.282$.

The total decay width of $X(4274)$ as the $3^{3}P_{1}$ state is lower than the 
experimental measurement performed by LHCb~\cite{Aaij:2016iza, Aaij:2016nsc}. 
It is worth to mention that the $X(4274)$ has been measured by the 
CDF~\cite{Aaltonen:2011at} and CMS~\cite{Chatrchyan:2013dma} Collaborations 
obtaining similar masses than the one of the LHCb but lower values of its total 
decay width: $32^{+22}_{-15}\pm8$ and $38^{+30}_{-15}\pm16$, respectively. These 
central values are in agreement with our theoretical prediction; in any case, 
the LHCb determination is fairly compatible with our figure. The information of 
the partial decay widths shown in Table~\ref{tab:chic13P} points out that the 
$DD^{\ast}$ and $D_{s}D_{s}^{\ast}$ decay channels are dominant with branching 
ratios of $\sim60\%$ and $\sim30\%$, respectively.

\begin{table}[!t]
\begin{center}
\begin{tabular}{lllrr}
\hline
\hline
State & $nL$  & Channel & $\Gamma$ (MeV) & ${\cal B}$ (\%) \\
\hline
$\chi_{c1}$ & $3P$ & $DD$ & $-$ & $-$ \\
            &      & $DD^{\ast}$ & $17.35$ & $58.24$ \\
            &      & $DD_{0}^{\ast}$ & $0.26$ & $0.88$ \\            
            &      & $D^{\ast}D^{\ast}$ & $0.43$ & $1.44$ \\
            &      & $D_{s}D_{s}$ & $-$ & $-$ \\
            &      & $D_{s}D_{s}^{\ast}$ & $8.49$ & $28.48$ \\
            &      & $D_{s}^{\ast}D_{s}^{\ast}$ & $3.26$ & $1.95$ \\
\multicolumn{2}{c}{$56\pm11^{+8}_{-11}$} & Total & $29.8$ & $100.00$ \\        
    
\hline
\hline
\end{tabular}
\caption{\label{tab:chic13P} Open-flavour strong decay widths, in MeV, and
branching fractions, in $\%$, of the $X(4274)$ meson with quantum numbers 
$nJ^{PC}=3\,1^{++}$. The experimental value of the total decay width is taken 
from Ref.~\cite{Aaij:2016iza, Aaij:2016nsc}.}
\end{center}
\end{table} 

\begin{table}[!t]
\begin{center}
\begin{tabular}{lllrr}
\hline
\hline
State & $nL$  & Channel & $\Gamma$ (MeV) & ${\cal B}$ (\%)\\
\hline
$\chi_{c0}$ & $4P$ & $DD$ & $13.27$ & $11.53$ \\
            &      & $DD^{\ast}$ & $-$ & $-$ \\
            &      & $DD_{0}^{\ast}$ & $-$ & $-$ \\
            &      & $DD_{1}$ & $19.50$ & $16.94$ \\
            &      & $DD_{1}^{\prime}$ & $27.23$ & $23.65$ \\
            &      & $DD_{2}^{\ast}$ & $-$ & $-$ \\
            &      & $D^{\ast}D^{\ast}$ & $2.19$ & $1.90$ \\            
            &      & $D^{\ast}D_{0}^{\ast}$ & $0.86$ & $0.75$ \\
            &      & $D^{\ast}D_{1}$ & $3.18$ & $2.76$ \\
            &      & $D^{\ast}D_{1}^{\prime}$ & $25.86$ & $22.47$ \\
            &      & $D^{\ast}D_{2}^{\ast}$ & $18.12$ & $15.74$ \\           
            &      & $D_{s}D_{s}$ & $0.06$ & $0.05$ \\
            &      & $D_{s}D_{s}^{\ast}$ & $-$ & $-$ \\
            &      & $D_{s}D_{s0}^{\ast}$ & $-$ & $-$ \\
            &      & $D_{s}D_{s1}(2460)$ & $0.74$ & $0.64$ \\
            &      & $D_{s}^{\ast}D_{s}^{\ast}$ & $3.76$ & $3.27$ \\
            &      & $D_{s}^{\ast}D_{s0}^{\ast}$ & $0.33$ & $0.29$ \\
\multicolumn{2}{c}{$92\pm21^{+21}_{-20}$} & Total & $115.11$ & $100.00$ \\ 
\hline
\hline
\end{tabular}
\caption{\label{tab:chic04P} Open-flavour strong decay widths, in MeV, and
branching fractions, in $\%$, of the $X(4500)$ meson with quantum numbers 
$nJ^{PC}=4\,0^{++}$. The experimental value of the total decay width is taken 
from Ref.~\cite{Aaij:2016iza, Aaij:2016nsc}.}
\end{center}
\end{table}

\begin{table}[!t]
\begin{center}
\begin{tabular}{lllrr}
\hline
\hline
State & $nL$  & Channel & $\Gamma$ (MeV) & ${\cal B}$ (\%)\\
\hline
$\chi_{c0}$ & $5P$ & $DD$ & $12.32$ & $10.10$ \\
            &      & $DD^{\ast}$ & $-$ & $-$ \\
            &      & $DD_{0}^{\ast}$ & $-$ & $-$ \\
            &      & $DD_{1}$ & $6.93$ & $5.68$ \\
            &      & $DD_{1}^{\prime}$ & $3.61$ & $2.96$ \\
            &      & $DD_{2}^{\ast}$ & $-$ & $-$ \\
            &      & $D^{\ast}D^{\ast}$ & $8.77$ & $7.19$ \\            
            &      & $D^{\ast}D_{0}^{\ast}$ & $5.69$ & $4.66$ \\
            &      & $D^{\ast}D_{1}$ & $2.32$ & $1.90$ \\
            &      & $D^{\ast}D_{1}^{\prime}$ & $20.39$ & $16.71$ \\
            &      & $D^{\ast}D_{2}^{\ast}$ & $56.22$ & $46.07$ \\           
            &      & $D_{s}D_{s}$ & $0.11$ & $0.09$ \\
            &      & $D_{s}D_{s}^{\ast}$ & $-$ & $-$ \\
            &      & $D_{s}D_{s0}^{\ast}$ & $-$ & $-$ \\
            &      & $D_{s}D_{s1}(2460)$ & $2.41$ & $1.98$ \\
            &      & $D_{s}D_{s1}(2536)$ & $0.26$ & $0.22$ \\
            &      & $D_{s}D_{s2}^{\ast}$ & $-$ & $-$ \\            
            &      & $D_{s}^{\ast}D_{s}^{\ast}$ & $1.36$ & $1.12$ \\
            &      & $D_{s}^{\ast}D_{s0}^{\ast}$ & $1.27$ & $1.04$ \\
            &      & $D_{s}^{\ast}D_{s1}(2460)$ & $0.29$ & $0.24$ \\
            &      & $D_{s}^{\ast}D_{s1}(2536)$ & $0.00$ & $0.00$ \\
            &      & $D_{s}^{\ast}D_{s2}^{\ast}$ & $0.03$ & $0.02$ \\
            &      & $D_{s0}^{\ast}D_{s0}^{\ast}$ & $0.03$ & $0.03$ \\
\multicolumn{2}{c}{$120\pm30^{+42}_{-33}$} & Total & $122.02$ & $100.00$ \\
\hline
\hline
\end{tabular}
\caption{\label{tab:chic05P} Open-flavour strong decay widths, in MeV, and
branching fractions, in $\%$, of the $X(4700)$ meson with quantum numbers 
$nJ^{PC}=5\,0^{++}$. The experimental value of the total decay width is taken 
from Ref.~\cite{Aaij:2016iza, Aaij:2016nsc}.}
\end{center}
\end{table}

One can see in Tables~\ref{tab:chic04P} and~\ref{tab:chic05P} that the 
predicted total decay widths for the $X(4500)$ and $X(4700)$ mesons as 
$J^{PC}=0^{++}$ $4P$ and $5P$ states are, within errors, in good agreement with 
the LHCb observations. Table~\ref{tab:chic04P} shows that the $DD$, 
$DD_{1}^{(\prime)}$, $D^{\ast}D_{1}^{\prime}$ and $D^{\ast}D_{2}^{\ast}$ decay 
channels are the most important for the $X(4500)$ meson with branching fractions 
ranging between $10\%$ and $25\%$. Table~\ref{tab:chic05P} shows that the 
$X(4700)$ decays around $50\%$ of the times into $D^{\ast}D_{2}^{\ast}$ final 
state. Traces in many other channels are found with partial decay widths of 
several MeV, the most important ones are its decays into $DD$ and 
$D^{\ast}D_{1}^{\prime}$ final states.


To gain some insight into the nature of the $X(4140)$, that does not appear as 
quark-antiquark state, and to see how the coupling with the open-flavour 
thresholds can modify the properties of the naive quark-antiquark states 
predicted above, we have performed a coupled-channel calculation including the 
$D^{\ast}D_{1}^{(\prime)}$, $D_{s}D_{s}$, $D_{s}^{\ast}D_{s}^{\ast}$ and 
$J/\psi\phi$ channels for the $J^{PC}=0^{++}$ sector; and the 
$D_{s}D_{s}^{\ast}$, $D_{s}^{\ast}D_{s}^{\ast}$ and $J/\psi\phi$ ones for the 
$J^{PC}=1^{++}$ sector. These are the allowed channels whose thresholds are in 
the region studied by the LHCb. It is important to remark here that, in 
principle, one should couple with the infinite number of meson-meson thresholds 
but it has been argued by many theorists~\cite{Swanson:2005rc, Barnes:2007xu} 
that the only relevant thresholds are those close to the naive states having the 
rest a little effect which can be absorbed in our quark model parameters.

Therefore, we assume now that the hadronic state can be described as 
\begin{equation} 
| \Psi \rangle = \sum_\alpha c_\alpha | \psi_\alpha \rangle
+ \sum_\beta \chi_\beta(P) |\phi_A \phi_B \beta \rangle,
\label{eq:funonda}
\end{equation}
where $|\psi_\alpha\rangle$ are $c\bar{c}$ eigenstates solution of the two-body 
problem, $\phi_{A}$ and $\phi_{B}$ are the two meson states with $\beta$ 
quantum numbers, and $\chi_\beta(P)$ is the relative wave function between the 
two mesons.

Two- and four-quark configurations are coupled using the same transition 
operator that has allowed us to compute the above open-flavour strong decays. 
This is because the coupling between the quark-antiquark and meson-meson 
sectors requires also the creation of a light quark 
pair~\cite{LeYaouanc:1972ae, LeYaouanc:1973xz}. We define the transition 
potential $h_{\beta \alpha}(P)$ within the $^{3}P_{0}$ 
model as~\cite{Kalashnikova:2005ui} 
\begin{equation}
\langle \phi_{A} \phi_{B} \beta | T | \psi_\alpha \rangle = P \, h_{\beta 
\alpha}(P) \,\delta^{(3)}(\vec P_{\rm cm})\,,
\label{eq:Vab}
\end{equation}
where $P$ denotes the relative momentum of the two-meson state.

Using Eq.~(\ref{eq:funonda}) and the transition potential in 
Eq.~(\ref{eq:Vab}), we arrive to the coupled equations
\begin{align}
&
c_\alpha M_\alpha +  \sum_\beta \int h_{\alpha\beta}(P) \chi_\beta(P)P^2 dP = E
c_\alpha\,, \label{ec:Ec-Res1} \\
&
\sum_{\beta}\int H_{\beta'\beta}(P',P)\chi_{\beta}(P) P^2 dP + \nonumber \\
&
\hspace{2.50cm} + \sum_\alpha h_{\beta'\alpha}(P') c_\alpha = E
\chi_{\beta'}(P') \,, \label{ec:Ec-Res2}
\end{align}
where $M_\alpha$ are the masses of the bare $c\bar{c}$ mesons and 
$H_{\beta'\beta}$ is the resonant group method (RGM) Hamiltonian for the 
two-meson states obtained from the $q\bar{q}$ interaction.

Solving the coupled-channel equations, Eqs.~(\ref{ec:Ec-Res1}) 
and~(\ref{ec:Ec-Res2}), as indicated in, e.g., Ref.~\cite{Ortega:2012rs}, we 
obtain the results shown in Tables~\ref{tab:r1} and~\ref{tab:r2} for the 
$J^{PC}=0^{++}$ channel and in Tables~\ref{tab:r3} and~\ref{tab:r4} for the 
$J^{PC}=1^{++}$ channel.

Table~\ref{tab:r1} shows that we obtain two states with $J^{PC}=0^{++}$ quantum 
numbers made by a $\sim\!\!50\%$ of $c\bar c$ component and by a similar amount 
of molecular components. Their masses are close to those associated with the 
bare $q\bar q$ $J^{PC}=0^{++}$ $4P$ and $5P$ states. Table~\ref{tab:r2} shows 
that the object with a mass of $4493.6\,{\rm MeV}$ is almost a pure $4P$ $c\bar 
c$ state whereas the object with a mass of $4674.1\,{\rm MeV}$ is almost a pure 
$5P$ $c\bar c$ state. Then, we first conclude that the net effect of coupling 
the thresholds to both naive quark-antiquark states is to modify the mass of 
the bare $c\bar c$ states in a modest amount. The second observation is that 
the total decay widths of these two states are significantly reduced. The new 
values, $79.2\,{\rm MeV}$ and $50.2\,{\rm MeV}$, are lower than the 
central ones reported by the LHCb but still within the experimental uncertainty 
interval.

In the coupled-channel calculation of the $J^{PC}=1^{++}$ channel, we include 
the $D_{s}D_{s}^{\ast}$, $D_{s}^{\ast}D_{s}^{\ast}$ and $J/\psi\phi$ thresholds. 
We found only one state with mass $4242.4\,{\rm MeV}$ and total decay width 
$25.9\,{\rm MeV}$. This state is made by $48.7\%$ of the $3P$ charmonium state 
and by $43.5\%$ of the $D_{s}D_{s}^{\ast}$ component (see Tables~\ref{tab:r3} 
and~\ref{tab:r4}). When coupling with thresholds, the modification in the mass 
and width is small. Our total decay width is still compatible with the LHCb 
result, but indicates that the lower CDF and CMS measurements are in better 
agreement with our prediction. We again show in Table~\ref{tab:r4} that the 
dominant charmonium component of such physical state is the $3P$ $c\bar c$ one.

\begin{table}[!t]
\begin{center}
\begin{tabular}{cccccccc}
\hline
\hline
Mass & Width & ${\cal P}_{c\bar c}$ & ${\cal P}_{D^{\ast}D_{1}}$ & ${\cal 
P}_{D^{\ast}D_{1}^{\prime}}$ & ${\cal P}_{D_{s}D_{s}}$ & ${\cal 
P}_{D_{s}^{\ast}D_{s}^{\ast}}$ & ${\cal P}_{J/\psi\phi}$ \\
\hline
$4493.6$ & $79.2$ & $57.2$ & $8.4$ & $33.1$ & $0.9$  & $0.4$ & $<0.1$ \\
$4674.1$ & $50.2$ & $47.6$ & $27.2$ & $21.0$ & $1.6$ & $2.6$ & $<0.1$ \\
\hline
\hline
\end{tabular}
\caption{\label{tab:r1} Mass, in MeV, total decay width, in MeV, and 
probability of each Fock component, in \%, for the $X(4500)$ and $X(4700)$ 
mesons. The calculated widths include both the contributions of the $c\bar c$ 
and molecular components. The results have been calculated in the 
coupled-channel quark model.}
\end{center}
\end{table}

\begin{table}[!t]
\begin{center}
\begin{tabular}{ccccccc}
\hline
\hline
Mass (MeV) & ${\cal P}_{c\bar c}$ & ${\cal P}_{(n<3)P}$ & ${\cal P}_{3P}$ & 
${\cal P}_{4P}$ & ${\cal P}_{5P}$ & ${\cal P}_{(n>5)P}$ \\
\hline
$4493.6$ & $57.2$ & $3.033$ & $11.332$ & $80.037$ & $5.573$ & $0.026$ \\
$4674.1$ & $47.6$ & $0.014$ & $0.001$ & $2.062$  & $97.071$ & $0.853$ \\
\hline
\hline
\end{tabular}
\caption{\label{tab:r2} Probabilities, in \%, of $nP$ $c\bar c$ bare states for 
the $X(4500)$ and $X(4700)$ mesons.}
\end{center}
\end{table}

\begin{table}[!t]
\begin{center}
\begin{tabular}{cccccc}
\hline
\hline
Mass & Width & ${\cal P}_{c\bar c}$ & ${\cal P}_{D_{s}D_{s}^{\ast}}$ & ${\cal 
P}_{D_{s}^{\ast}D_{s}^{\ast}}$ & ${\cal P}_{J/\psi\phi}$ \\
\hline
$4242.4$ & $25.9$ & $48.7$ & $43.5$ & $5.0$ & $2.7$ \\
\hline
\hline
\end{tabular}
\caption{\label{tab:r3} Mass, in MeV, total decay width, in MeV, and 
probability of each Fock component, in \%, for the $X(4274)$ meson. The 
calculated widths include both the contributions of the $c\bar c$ 
and molecular components. The results have been calculated in the 
coupled-channel quark model.}
\end{center}
\end{table}
\begin{table}[h]
\begin{center}
\begin{tabular}{ccccccc}
\hline
\hline
Mass (MeV) & ${\cal P}_{c\bar c}$ & ${\cal P}_{1P}$ & ${\cal P}_{2P}$ & 
${\cal P}_{3P}$ & ${\cal P}_{4P}$ & ${\cal P}_{(n>4)P}$ \\
\hline
$4242.4$ & $48.7$& $0.000$ & $0.370$  & $99.037$ & $0.488$ & $0.105$ \\
\hline
\hline
\end{tabular}
\caption{\label{tab:r4} Probabilities, in \%, of $nP$ $c\bar c$ bare states for 
the $X(4274)$ meson.}
\end{center}
\end{table}


\begin{figure}
\begin{center}
\includegraphics[width=0.45\textwidth]{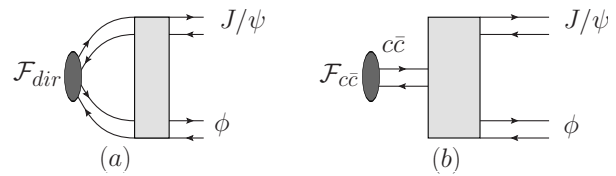} 
\end{center}
\caption{\label{fig:f1} Diagrams of the two possible production mechanisms for 
the $J/\psi\phi$ channel: Direct production through a point-like source (a) or 
production through intermediate $c\bar c$ states (b).}
\end{figure}

As we do not find any signal for the $X(4140)$, neither bound nor virtual, we 
analyze the line shape of the $J/\psi\phi$ channel as an attempt to explain 
the $X(4140)$ as a simple threshold cusp.

We evaluate the production of $J/\psi\phi$ pairs via two main mechanisms: (i) 
the direct generation of $J/\psi$ and $\phi$ mesons from a point-like source 
and (ii) the production via intermediate $c\bar c$ states. Therefore, the 
line-shape is given by
\begin{equation}
\label{eq:LineShape}
\frac{dB(J/\psi\phi)}{dE} = {\cal B} \, k \, \left[ \left|{\cal M}_{\rm point} 
\right|^{2} + \left|{\cal M}_{c\bar c}\right|^{2} \right] \Theta(E) \,,
\end{equation}
with $k$ the on-shell momentum. The ${\cal M}_{\rm point}$ is the direct 
production of $J/\psi\phi$ given by (Fig.~\ref{fig:f1}a):
\begin{equation}
\begin{split}
&
\mathcal{M}_{\rm point}^\beta(E) = \mathcal{F}_{\rm point} \times \\
&
\hspace*{0.70cm} \times \left(1- \sum_{\beta'}\int dP\, 
T^{\beta\beta'}(E;k,P)\frac{2\mu P^2}{P^2-k^2}\right)_{\rm on-shell} \,,
\end{split}
\end{equation}
where ${\cal F}_{\rm point}$ is the production amplitude from a point-like 
source.

In addition, ${\cal M}_{c\bar c}$ is the production via $c\bar c$ states 
(Fig.~\ref{fig:f1}b), which can be expressed as
\begin{equation} 
\mathcal{M}_{c\bar c}^\beta = -{\cal F}_{c\bar c} \sum_{\alpha\alpha'} 
\Phi_{\alpha'\beta}(E;k) \, \Delta_{\alpha'\alpha}(E)^{-1} \,, 
\label{eq:Mmesonico}
\end{equation}
where ${\cal F}_{c\bar c}$ is the production amplitude from a $c\bar c$ state, 
$\Phi$ is the $^3P_0$ vertex dressed by the RGM interaction and $\Delta$ 
is the complete propagator (see Ref.~\cite{Ortega:2012rs} for details).

\begin{figure}[!t]
\begin{center}
\epsfig{figure=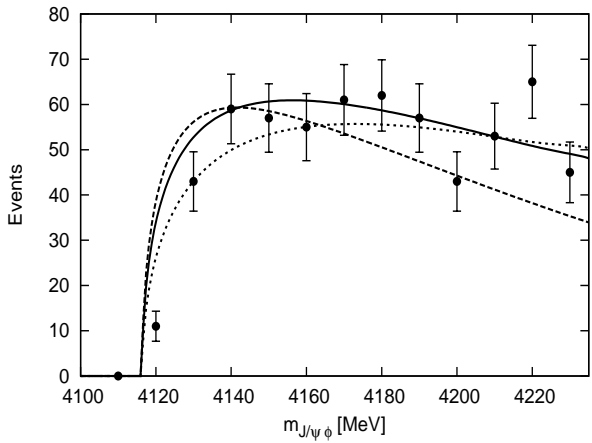,height=0.35\textheight,width=0.35\textwidth,
angle=-90}
\caption{\label{fig:f2} Line-shape prediction of the $J/\psi\phi$ channel. The 
solid curve is the result of Eq.~\eqref{eq:LineShape}. The other two curves 
show the production of $J/\psi\phi$ pairs via two main mechanisms: the direct 
generation of $J/\psi$ and $\phi$ mesons from a point-like source (dotted 
curve) and the production via intermediate $c\bar c$ states (dashed curve). 
Note that the normalization ${\cal B}$ in Eq.~\eqref{eq:LineShape} has been 
fitted to the data for each curve.}
\end{center}
\end{figure}

Figure~\ref{fig:f2} compares our result with that reported by the LHCb 
Collaboration in the $B^+\to J/\psi\phi K^+$ decays. The rapid increasing 
observed in the data near the $J/\psi\phi$ threshold corresponds with a bump in 
the theoretical result just above such threshold. This cusp is too wide to be 
produced by a bound or virtual state below the $J/\psi\phi$ threshold. If one 
fixes, attending to the experimental data, the normalization ${\cal B}$ in 
Eq.~\eqref{eq:LineShape} to each contribution separately, one realizes that the 
production via intermediate $c\bar c$ states is dominant at low values of the 
invariant mass, which is reasonable as the point-like production should be 
suppressed because the number of particles to be created is twice; whereas, at 
high energies, it is the direct generation of $J/\psi$ and $\phi$ mesons from a 
point-like source which drives the production.

As a summary, we have analysed in a coupled-channel quark model the 
$J/\psi\phi$ structures reported by the LHCb Collaboration in the amplitude 
analysis of the $B^{+}\to J/\psi\phi K^+$ decays. Three of them, namely 
$X(4274)$, $X(4500)$ and $X(4700)$ are consistent with bare quark-antiquark 
states with quantum numbers, respectively, $J^{PC}=1^{++} (3P)$, $J^{PC}=0^{++} 
(4P)$ and $J^{PC}=0^{++} (5P)$. The agreement between theory and experiment is 
due in part by our particular choice of the confining potential: 
screened-linear. Partial and total decay widths of the $X(4274)$, $X(4500)$ and 
$X(4700)$ mesons have been also computed using a version of the $^{3}P_{0}$ 
decay model in which its only parameter has been constrained before in other 
meson sectors. The coupling of the naive quark-antiquark states with the near 
meson-meson thresholds modifies slightly their properties but does not generate 
new resonances.

In the $1^{++}$ sector we do not find any pole in the mass region of the 
$X(4140)$. However, the scattering amplitude shows a bump just above the 
$J/\psi\phi$ threshold which reproduces the rapid increasing of the experimental 
data. Therefore, the structure showed by this data around $4140\,{\rm MeV}$ 
should be interpreted as a cusp due to the presence of the $D_{s}D_{s}^{\ast}$ 
threshold. The residual $D_{s}-D_{s}^{\ast}$ interaction is too weak to 
develop a bound or virtual state.

The experimental results constitute a prominent example of the interplay 
between quark and meson degrees of freedom in near open-flavoured threshold 
regions. Depending on the dynamics of the system, the presence of these 
thresholds can generate new states, simply renormalizes the masses of the bare 
$q\bar q$ states or produces cusp effects at threshold when the interactions 
are not strong enough to produce bound states.

Finally, these results also reinforce the validity of the constituent quark 
model to qualitatively describe the phenomenology of the excited heavy quark 
meson states and get insights on the dynamics that leads their formation.


\begin{acknowledgments}
This work has been partially funded by Ministerio de Ciencia y Tecnolog\'\i a 
under Contract no. FPA2013-47443-C2-2-P, by the Spanish Excellence Network 
on Hadronic Physics FIS2014-57026-REDT, and by the Junta de Castilla y Le\'on 
under Contract no. SA041U16. P. G. Ortega acknowledges the financial support of 
the Spanish Ministerio de Econom\'ia y Competitividad and European FEDER funds 
under the contracts FIS2014-51948-C2-1-P. J. Segovia acknowledges the financial 
support from Alexander von Humboldt Foundation.
\end{acknowledgments}


\bibliography{NotAllResonancesXYZ}

\end{document}